\documentclass[aps,prl,twocolumn,showpacs,superscriptaddress,groupedaddress]{revtex4-1}

\usepackage{graphicx}

\usepackage[T1]{fontenc}
\usepackage[utf8]{inputenc}

\begin{document}
\title{First-principles study of point defects at semicoherent interface}

\author{E. Metsanurk} \affiliation{Department of Physics and Astronomy, Uppsala University, Box 516, S-75120 Uppsala, Sweden}
\author{A. Caro} \affiliation{Los Alamos National Laboratory, Los Alamos, New Mexico 87545, USA}
\author{A. Tamm} \affiliation{Intelligent Materials and Systems Laboratory, Institute of Technology, University of Tartu, Tartu, Estonia}
\author{A. Aabloo} \affiliation{Intelligent Materials and Systems Laboratory, Institute of Technology, University of Tartu, Tartu, Estonia}
\author{M. Klintenberg} \affiliation{Department of Physics and Astronomy, Uppsala University, Box 516, S-75120 Uppsala, Sweden}

\date{\today}

\begin{abstract}
Modeling semicoherent metal-metal interfaces has so far been performed 
using atomistic simulations based on semiempirical interatomic 
potentials. We demonstrate through more precise ab-initio calculations 
that key conclusions drawn from previous studies do not conform with 
the new results which show that single point defects do not delocalize 
near the interfacial plane, but remain compact. We give a simple 
qualitative explanation for the difference in predicted results that 
can be traced back to limited transferability of empirical potentials.
\end{abstract}

\maketitle

Nanostructured metallic multilayer composites (NMMC) are known to have 
superior mechanical properties compared to standard coarse grained 
metals \cite{Beyerlein2012} along with the ability to efficiently 
self-heal radiation damage \cite{Misra2007,Gao2011}. The latter is 
crucial for the material to inhibit creep and swelling in harsh 
environments. In order to utilize these materials in most effective 
manner, it is necessary to understand the underlying mechanisms leading 
to the aforesaid advantages. One route to accomplish that is via 
experiments where state of the art work has reached to a point where it 
is possible to engineer bulk nanostructured bi-metal multilayers while 
having control over structural features at atomic level which can 
considerably alter the mechanical properties and thermal stability of 
these materials \cite{Zheng2014}.

Another way to gain insight into the way these materials behave during 
subjection to extreme mechanical or radiation environments is by 
theoretical means via computational methods. Since the time and length 
scales these processes cover are exceptionally widely spread, starting 
from attoseconds and picometers for electronic effects and going up to 
meters and years in continuum mechanics, there is no single equation or 
model that can currently cover all of this complexity. Therefore only a 
multi-scale modeling approach allows to eventually predict the 
properties and design optimal NMMCs for future industrial and energy 
technology applications.

This study concentrates on the atomic level part of the multi-scale 
method, where there has been considerable effort to model the structure 
and behavior of NMMCs in order to better understand the traits leading 
to the high tolerance to radiation damage. While the effects caused by 
irradiation are essentially macroscopic, they are still governed by 
changes in atomic level that can be traced back to single point 
defects. Therefore efforts have been directed to identifying possible 
lowest energy structures for the undamaged interfaces as well as 
describing the point defect properties such as configurations, 
formation energies, migration barriers and mechanisms near the 
interfacial plane. Previous studies have shown that the interface does 
not support conventional point defects, that is vacancies and 
interstitial atoms, but instead pairs of extended jogs will form. Those 
delocalized defects have been shown to exhibit low formation energies 
and interaction through long-range forces \cite {Demkowicz2008a}. This 
will also result in more complex migration pathways and recombination 
mechanisms than in bulk material \cite 
{Kolluri2010,Kolluri2012,Liu2012}.

The studies described above were performed using atomistic modeling 
based on classical molecular dynamics and empirical interatomic 
potentials which have a crucial role in determining the outcome of the 
simulations. Fitting empirical models to ab-initio or experimental data 
always presents the challenge of obtaining good transferability to the 
problem under study which often explores regions of phase space not 
used in the fitting process. For metallic systems the most widely 
utilized model is the embedded-atom method (EAM) developed by Daw and 
Baskes \cite {Daw1983,Daw1984}. Although there are alternatives, 
arguably having greater accuracy, it is still relevant because of its 
simplicity and computational scalability while providing relatively 
accurate description, especially for FCC metals.

There are two EAM potentials available for copper-niobium system which were 
fitted using two different methods. First by Demkowicz et 
al \cite {Demkowicz2009} (hereafter EAM1) uses modified Morse 
function for Cu-Nb interaction and has been fitted to dilute 
enthalpies of mixing and bulk modulus and lattice constant of 
hypothetical CuNb alloy in B2 structure. The second one by 
Zhang et al \cite{Zhang2013} (EAM2) uses more flexible 
polynomial-like function and is fitted to enthalpies of mixing of 
special quasi-random structures over the whole composition range 
with the aim of correctly reproducing experimental thermodynamics for 
the system.

In this work we show, that EAM1 and EAM2 give markedly different 
results for both the structure and energetics of point defects near the 
interface. Then we propose a solution to this discrepancy by relaxing 
the structures predicted by aforementioned two potentials using 
density-functional theory (DFT) calculations which essentially do not 
rely on empirical parameters thereby producing more accurate results. 
We then propose an explanation why some interatomic potentials might 
lead to erroneous characterization of the interface and how to possibly 
prevent this in future works.

All DFT calculations were done using plane-wave pseudopotential code 
Vienna Ab initio Simulation Package (VASP) \cite
{Kresse1993,Kresse1994a,Kresse1996,Kresse1996a} with supplied PAW 
pseudopotentials \cite{Kresse1999} and GGA-PBE approximation \cite
{Perdew1996,Perdew1997}. For niobium the semi core $p$ states were 
treated as valence. The cutoff energy for the plane waves was 
$273.214$~eV. Single k-point ($\Gamma$-point) was used and smearing 
was handled by 1st order Methfessel-Paxton scheme \cite
{Methfessel1989} with width of $\sigma=0.01$~eV. Atomistic simulations 
were performed with classical molecular dynamics code LAMMPS \cite
{Plimpton19951}.

\begin{figure}
\includegraphics[scale=0.4]{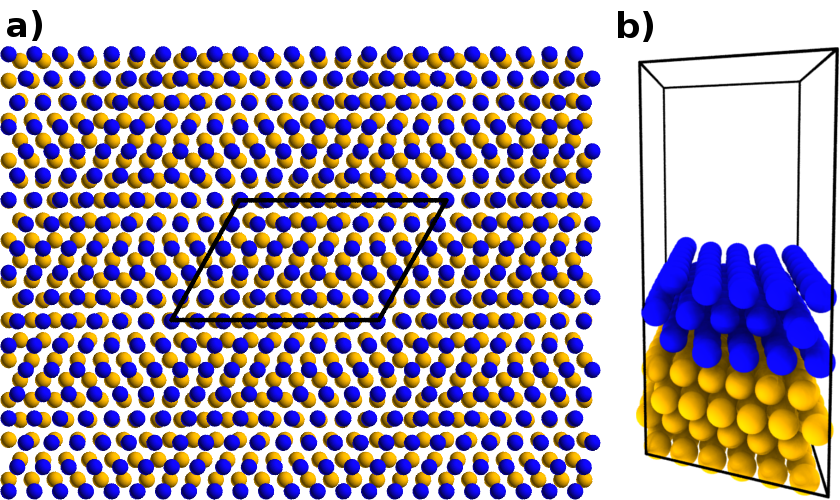}
\caption{\label{fig:quasiunit} Constructing small quasi-unit cell for 
Cu/Nb interface in Kurdjumov-Sachs orientation. Blue spheres represent 
niobium and yellow spheres copper atoms. (a) The unit cell is ``cut 
out'' from larger structure and has approximately same periodicity as 
the misfit dislocation intersections, that is the areas where copper 
and niobium atom positions coincide in x- and y-directions. (b) A 3D 
view of the simulation cell, where each layer consists of either 54 Cu or 40 
Nb atoms. Total number of atoms is 336.}
\end{figure}

The structure of the interface can be described by specifying the 
orientation of the surface normal to the interfacial plane and two 
parallel directions, one for each surface, that will be parallel when 
the interface is formed. It has been shown experimentally that 
copper-niobium interface forms predominantly in Kurdjumov-Sachs 
orientation \cite{Anderson2003}. In general calculating the energies of 
such structures using DFT is a complex task solely because of the 
number of atoms needed, and hence the required computational effort, to 
retain characteristic features and periodicities of the interface. The 
periodicity of the interface is defined by the locations of misfit 
dislocation intersections (MDIs), that is the areas where the atoms on 
each side of the interface overlap \cite{Demkowicz2008}. In case of 
copper and niobium in KS orientation the distances between the MDIs are 
relatively small enabling this specific interface to be modeled using a 
quasi-unit cell appropriately sized for DFT. This cell is an 
approximation and not a true unit cell for the larger system since 
albeit similar, the local environments around the MDIs are not 
equivalent. The setup is illustrated in Fig. \ref {fig:quasiunit}.

In order to keep the calculations computationally feasible the number 
of atoms in the cell perpendicular to the interface must be limited. 
This results in two choices, either make the simulation box periodic or 
add vacuum in this direction. Former corresponds to having infinite 
number of thin alternating copper and niobium layers and latter to 
single interface and two free surfaces. We opted for having $1.65$~nm 
layer of vacuum between the free surfaces. Unit cell vectors (in nm) 
for the resulting unit cell are $a_x = (2.3, 0, 0)$, $a_y = (0.75, 
1.33, 0)$ and $a_z = (0,0,3.0)$ and it consists of 216 Cu and 120 Nb 
atoms. We checked for possible errors by doubling the number of layers 
and calculating the structure, which did not change, and formation 
energy of vacancy, which reduced by 20 meV. While constructing a small 
unit cell using the method described above gives an appropriate 
representation of the undamaged interface, there is still the problem 
of finding ground state configurations of point defects. Therefore we 
calculated candidate structures using both EAM1 and EAM2 with molecular 
dynamics and then relaxed these using DFT. The ratio of lattice 
constants of Cu and Nb calculated using DFT and MD differ by less than
$0.1 \%$ thereby making this method valid.

\begin{figure}
\includegraphics[scale=0.3]{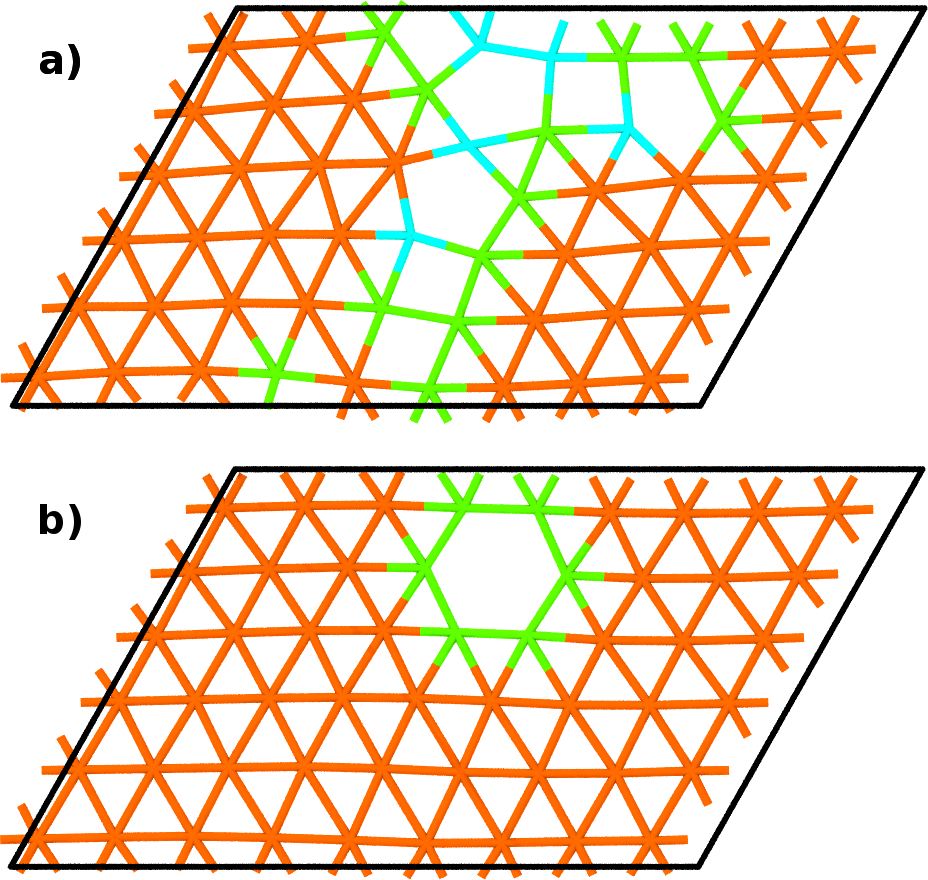}
\caption{\label{fig:dftrelax} Top view of interfacial copper layer with 
single vacancy at the interface before and after relaxing using DFT. 
(a) The structure was obtained by relaxing the interface with EAM1 
which results in delocalization of the defect and formation of 4- and 
5-atom rings. (b) Relaxing in DFT yields single compact vacancy. Lines 
and colors represent the in-plane bonds and 
coordination numbers respectively with orange being 6, green 5, 
cyan 4. The position of vacancy is at the MDI, but is shifted
due to layers moving with respect to each other in simulation. }
\end{figure}

In order to a) check whether the vacuum layer or distortion of the true 
periodicity has any effect on the structure of the interface, b) to get 
input structures for DFT calculations and c) to assess whether relaxing 
the box has important effect on the outcome of the results, we first 
performed different molecular dynamics simulations with both EAM1 and 
EAM2. First the initial system was quenched from 600 K to 0 K followed 
by the energy minimization using conjugate gradient method. Next the 
copper atom with highest potential energy was removed and the process 
was repeated. A typical final structure as predicted by EAM1 is shown 
in Fig. \ref{fig:dftrelax}a and has the same 4- and 5-atom rings as 
found in previous works while using EAM2 a compact single vacancy is 
formed. Relaxing the simulation box using EAM2 has minuscule effect on 
the formation energies while using EAM1 results in somewhat smaller 
energies. In either case the structural features are not affected.

\begin{figure}
\includegraphics[scale=0.3]{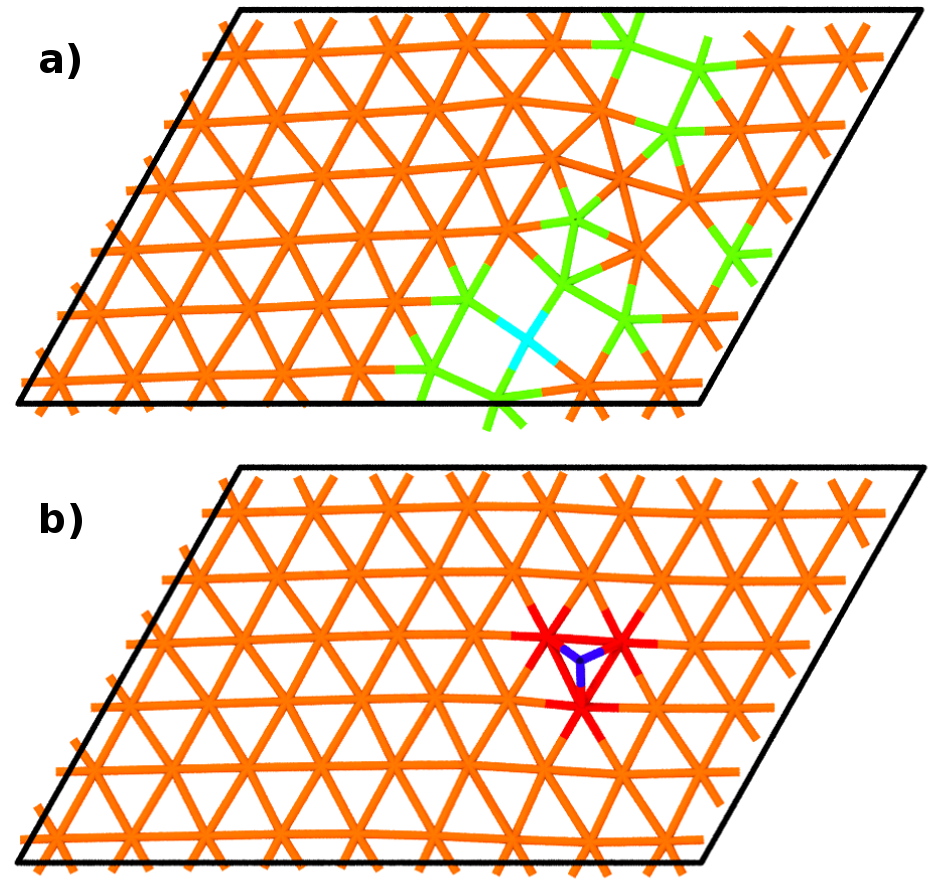}
\caption{\label{fig:dftrelax_int} Top view of interfacial copper 
layer with added interstitial atom before and after relaxing using DFT. 
(a) The structure was obtained by relaxing the interface with EAM1 
potential which results in delocalization of the defect and 
formation of 4-atom rings. (b) After relaxing in DFT a single 
interstitial between the interfacial copper and niobium layers will emerge. 
Lines and colors represent in-plane bonds and coordination numbers
respectively, with red being 7, orange 6, green 5 and cyan 4 and blue 3.}
\end{figure}

Same process was carried out with single interstitial copper atom which 
was inserted into the interface after initial energy minimization next 
to the MDI. Again using EAM1 results in delocalization of the defect 
(on Fig. \ref{fig:dftrelax_int}a) while EAM2 produces clear 
interstitial which resides between the copper and niobium layers.

Next all four structures were relaxed using DFT. The energy 
minimization was done using conjugate gradient method until the maximum 
force on an atom was less than 0.1 eV/\AA~while the energy difference 
between two sequential minimization steps was below 0.1 meV. The box 
size was kept constant for consistency and lower computational cost 
based on the fact that no change in structure and only negligible effect 
on formation energies was observed in constant pressure molecular 
dynamics runs using either EAM potential. Resulting structures are 
depicted on Fig. \ref {fig:dftrelax}a and Fig. \ref{fig:dftrelax_int}a 
for the vacancy and interstitial respectively. The final structures are 
very similar to the ones obtained with EAM2 potential, that is no 
delocalization happens and compact vacancy or interstitial is formed.

The formation energies for copper vacancies and interstitials at the 
interface calculated with the two potentials and DFT are listed in 
Table \ref{tab:energies}. It must be noted though, that the values 
cannot be directly compared. The reason for that is the difference in 
defect energies of pure copper which will carry over to the formation 
energies of defects near the interface. Similarly, these values cannot 
be compared to the ones calculated using larger unit cell, firstly 
because of possible defect-defect interaction in neighboring cells due 
to periodicity and secondly due to the probable errors introduced by 
approximating the large cell using a smaller one. Moreover, while the 
defect delocalization predicted by EAM1 can result in diverse 
final structures with different energies as reported in Ref. 
\cite{Demkowicz2008a}, a single point defect has a well-defined 
formation energy.

Our results demonstrate that different empirical potentials can lead to 
contrasting results when the structures studied are substantially 
different from those used for the fitting procedure. The result of the 
fitting of alloy properties in case of EAM is a single function 
relating distance between two different species to the energy. Since 
the data used to fit EAM1 depends only on a small discrete set of 
distances, the energy function is also well defined only at these 
points. At the same time the energy of a semicoherent interface 
contains distances of nearly continuous spectrum which means that when 
using this method the energy of the interface and thus the formation 
energies of defects can take nearly arbitrary values limited only by 
the chosen functional form.

The same method of fitting as described above has been also used to 
``tune'' the potentials to yield different enthalpies of mixing and to 
show how this would affect the behavior of vacancies and interstitials 
\cite {Liu2010,Kolluri2013}. While this method is a nice example of the 
ability of simulations to investigate scenarios which are impossible to 
achieve in experiments, having too small fitting database (too few 
distances in case of EAM) might again interfere with other physical 
quantities which can lead to incorrect conclusions.

\begin{figure}
\includegraphics[scale=0.3]{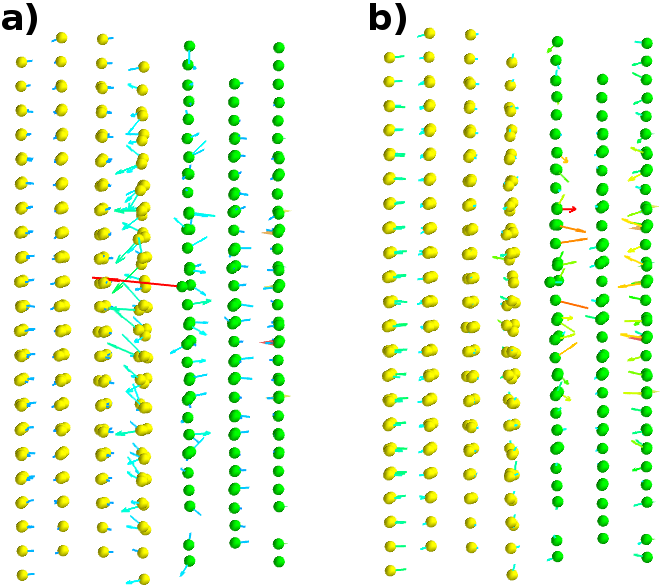}
\caption{\label{fig:mdforces} Difference between DFT and MD forces (in 
arb. units) on atoms for DFT-relaxed interface structure with single 
copper vacancy. (a) and (b) show the forces obtained using EAM1 and 
EAM2 respectively. EAM1 produces significantly larger forces on copper 
atoms near the interfacial plane compared to those of DFT and EAM2. 
Copper atoms - yellow, niobium atoms - green.}
\end{figure}

A method has been proposed by Ercolessi and Adams in which instead of 
energies, forces are fitted to reproduce those obtained from 
first-principles calculations. This has been shown to lead to greater 
accuracy and transferability of the potentials \cite {Ercolessi1994}. 
However, increasing the fitting database, as it is in the case of EAM2, 
can also result in significant improvement. Figure \ref{fig:mdforces} 
shows the forces on atoms for the relaxed DFT structure with one 
vacancy calculated using both EAM1 and EAM2. The forces predicted by 
latter differ predominantly at free surfaces and for niobium atoms. It 
has been shown that it is quite hard or even impossible to accurately 
reproduce DFT forces for niobium using EAM \cite{Fellinger2010} so this 
behavior is expected. With EAM1 the difference between DFT and MD 
forces for the first copper layer (where the vacancy is located) is 
much larger which leads to reconstruction of the layer and 
delocalization of the vacancy.

\begin{table}
\caption{\label{tab:energies}Formation energies of copper vacancies 
and interstitials at the interface in eV. EAM1 tends to underestimate the 
formation energies while EAM2 overestimates these. Values in 
parentheses represent same energies in bulk copper. It must be 
noted, that a) since the energies are calculated using relatively 
small number of atoms, they do not truly represent formation 
energies in dilute limit and b) care must be taken comparing DFT and 
MD values since the energies in pure copper already differ.}
\begin{ruledtabular}
\begin{tabular}{lccr}
Structure &EAM1\footnote{Potential by Demkowicz et al \cite{Demkowicz2009}}&EAM2\footnote{Potential by Zhang et al \cite{Zhang2013}}& DFT\footnote{This work}\\
\hline
Vacancy & 0.18 (1.26) & 0.72 (1.27) & 0.38 (1.17) \\
Interstitial & 1.07 (3.27) & 1.52 (3.09) & 1.13 (3.86) \\
\end{tabular}
\end{ruledtabular}
\end{table}

To summarize, we have shown that single compact point defects can exist 
at semicoherent metal-metal interfaces without any delocalization 
contrary to results of previous studies. This could necessitate further 
investigation of defect migration, clustering and recombination. In 
addition, we have provided an explanation for the discrepancy between 
earlier atomistic studies and this work which can be attributed to 
different fitting strategies and intrinsic transferability limitations 
of empirical potentials.

The research leading to these results has received funding from the 
European Union Seventh Framework Programme (FP7/2007-2013) under grant 
agreement n$^\circ$~263273. Computational resources were provided by 
Swedish National Infrastructure for Computing. Work of A. Caro 
performed at the Center for Materials at Irradiation and Mechanical 
Extremes, an Energy Frontier Research Center funded by the U.S. 
Department of Energy (Award No. 2008LANL1026) at Los Alamos National 
Laboratory.

\bibliography{cunb_dft,vasp_cite}{}

\begin{thebibliography}{27}%
\makeatletter
\providecommand \@ifxundefined [1]{%
 \@ifx{#1\undefined}
}%
\providecommand \@ifnum [1]{%
 \ifnum #1\expandafter \@firstoftwo
 \else \expandafter \@secondoftwo
 \fi
}%
\providecommand \@ifx [1]{%
 \ifx #1\expandafter \@firstoftwo
 \else \expandafter \@secondoftwo
 \fi
}%
\providecommand \natexlab [1]{#1}%
\providecommand \enquote  [1]{``#1''}%
\providecommand \bibnamefont  [1]{#1}%
\providecommand \bibfnamefont [1]{#1}%
\providecommand \citenamefont [1]{#1}%
\providecommand \href@noop [0]{\@secondoftwo}%
\providecommand \href [0]{\begingroup \@sanitize@url \@href}%
\providecommand \@href[1]{\@@startlink{#1}\@@href}%
\providecommand \@@href[1]{\endgroup#1\@@endlink}%
\providecommand \@sanitize@url [0]{\catcode `\\12\catcode `\$12\catcode
  `\&12\catcode `\#12\catcode `\^12\catcode `\_12\catcode `\%12\relax}%
\providecommand \@@startlink[1]{}%
\providecommand \@@endlink[0]{}%
\providecommand \url  [0]{\begingroup\@sanitize@url \@url }%
\providecommand \@url [1]{\endgroup\@href {#1}{\urlprefix }}%
\providecommand \urlprefix  [0]{URL }%
\providecommand \Eprint [0]{\href }%
\providecommand \doibase [0]{http://dx.doi.org/}%
\providecommand \selectlanguage [0]{\@gobble}%
\providecommand \bibinfo  [0]{\@secondoftwo}%
\providecommand \bibfield  [0]{\@secondoftwo}%
\providecommand \translation [1]{[#1]}%
\providecommand \BibitemOpen [0]{}%
\providecommand \bibitemStop [0]{}%
\providecommand \bibitemNoStop [0]{.\EOS\space}%
\providecommand \EOS [0]{\spacefactor3000\relax}%
\providecommand \BibitemShut  [1]{\csname bibitem#1\endcsname}%
\let\auto@bib@innerbib\@empty
\bibitem [{\citenamefont {Beyerlein}\ \emph {et~al.}(2012)\citenamefont
  {Beyerlein}, \citenamefont {Mara}, \citenamefont {Wang}, \citenamefont
  {Carpenter}, \citenamefont {Zheng}, \citenamefont {Han}, \citenamefont
  {Zhang}, \citenamefont {Kang}, \citenamefont {Nizolek},\ and\ \citenamefont
  {Pollock}}]{Beyerlein2012}%
  \BibitemOpen
  \bibfield  {author} {\bibinfo {author} {\bibfnamefont {I.~J.}\ \bibnamefont
  {Beyerlein}}, \bibinfo {author} {\bibfnamefont {N.~a.}\ \bibnamefont {Mara}},
  \bibinfo {author} {\bibfnamefont {J.}~\bibnamefont {Wang}}, \bibinfo {author}
  {\bibfnamefont {J.~S.}\ \bibnamefont {Carpenter}}, \bibinfo {author}
  {\bibfnamefont {S.~J.}\ \bibnamefont {Zheng}}, \bibinfo {author}
  {\bibfnamefont {W.~Z.}\ \bibnamefont {Han}}, \bibinfo {author} {\bibfnamefont
  {R.~F.}\ \bibnamefont {Zhang}}, \bibinfo {author} {\bibfnamefont
  {K.}~\bibnamefont {Kang}}, \bibinfo {author} {\bibfnamefont {T.}~\bibnamefont
  {Nizolek}}, \ and\ \bibinfo {author} {\bibfnamefont {T.~M.}\ \bibnamefont
  {Pollock}},\ }\href {\doibase 10.1007/s11837-012-0431-0} {\bibfield
  {journal} {\bibinfo  {journal} {Jom}\ }\textbf {\bibinfo {volume} {64}},\
  \bibinfo {pages} {1192} (\bibinfo {year} {2012})}\BibitemShut {NoStop}%
\bibitem [{\citenamefont {Misra}\ \emph {et~al.}(2007)\citenamefont {Misra},
  \citenamefont {Demkowicz}, \citenamefont {Zhang},\ and\ \citenamefont
  {Hoagland}}]{Misra2007}%
  \BibitemOpen
  \bibfield  {author} {\bibinfo {author} {\bibfnamefont {A.}~\bibnamefont
  {Misra}}, \bibinfo {author} {\bibfnamefont {M.~J.}\ \bibnamefont
  {Demkowicz}}, \bibinfo {author} {\bibfnamefont {X.}~\bibnamefont {Zhang}}, \
  and\ \bibinfo {author} {\bibfnamefont {R.~G.}\ \bibnamefont {Hoagland}},\
  }\href {\doibase 10.1007/s11837-007-0120-6} {\bibfield  {journal} {\bibinfo
  {journal} {JOM}\ }\textbf {\bibinfo {volume} {59}},\ \bibinfo {pages} {62}
  (\bibinfo {year} {2007})}\BibitemShut {NoStop}%
\bibitem [{\citenamefont {Gao}\ \emph {et~al.}(2011)\citenamefont {Gao},
  \citenamefont {Yang}, \citenamefont {Xue}, \citenamefont {Yan}, \citenamefont
  {Zhou}, \citenamefont {Wang}, \citenamefont {Kwok}, \citenamefont {Chu},\
  and\ \citenamefont {Zhang}}]{Gao2011}%
  \BibitemOpen
  \bibfield  {author} {\bibinfo {author} {\bibfnamefont {Y.}~\bibnamefont
  {Gao}}, \bibinfo {author} {\bibfnamefont {T.}~\bibnamefont {Yang}}, \bibinfo
  {author} {\bibfnamefont {J.}~\bibnamefont {Xue}}, \bibinfo {author}
  {\bibfnamefont {S.}~\bibnamefont {Yan}}, \bibinfo {author} {\bibfnamefont
  {S.}~\bibnamefont {Zhou}}, \bibinfo {author} {\bibfnamefont {Y.}~\bibnamefont
  {Wang}}, \bibinfo {author} {\bibfnamefont {D.~T.}\ \bibnamefont {Kwok}},
  \bibinfo {author} {\bibfnamefont {P.~K.}\ \bibnamefont {Chu}}, \ and\
  \bibinfo {author} {\bibfnamefont {Y.}~\bibnamefont {Zhang}},\ }\href
  {\doibase 10.1016/j.jnucmat.2011.03.030} {\bibfield  {journal} {\bibinfo
  {journal} {Journal of Nuclear Materials}\ }\textbf {\bibinfo {volume}
  {413}},\ \bibinfo {pages} {11} (\bibinfo {year} {2011})}\BibitemShut
  {NoStop}%
\bibitem [{\citenamefont {Zheng}\ \emph {et~al.}(2014)\citenamefont {Zheng},
  \citenamefont {Carpenter}, \citenamefont {McCabe}, \citenamefont
  {Beyerlein},\ and\ \citenamefont {Mara}}]{Zheng2014}%
  \BibitemOpen
  \bibfield  {author} {\bibinfo {author} {\bibfnamefont {S.}~\bibnamefont
  {Zheng}}, \bibinfo {author} {\bibfnamefont {J.~S.}\ \bibnamefont
  {Carpenter}}, \bibinfo {author} {\bibfnamefont {R.~J.}\ \bibnamefont
  {McCabe}}, \bibinfo {author} {\bibfnamefont {I.~J.}\ \bibnamefont
  {Beyerlein}}, \ and\ \bibinfo {author} {\bibfnamefont {N.~a.}\ \bibnamefont
  {Mara}},\ }\href {\doibase 10.1038/srep04226} {\bibfield  {journal} {\bibinfo
   {journal} {Scientific reports}\ }\textbf {\bibinfo {volume} {4}},\ \bibinfo
  {pages} {4226} (\bibinfo {year} {2014})}\BibitemShut {NoStop}%
\bibitem [{\citenamefont {Demkowicz}\ \emph {et~al.}(2008)\citenamefont
  {Demkowicz}, \citenamefont {Hoagland},\ and\ \citenamefont
  {Hirth}}]{Demkowicz2008a}%
  \BibitemOpen
  \bibfield  {author} {\bibinfo {author} {\bibfnamefont {M.~J.}\ \bibnamefont
  {Demkowicz}}, \bibinfo {author} {\bibfnamefont {R.G.}~\bibnamefont {Hoagland}},
  \ and\ \bibinfo {author} {\bibfnamefont {J.P.}~\bibnamefont {Hirth}},\ }\href
  {\doibase 10.1103/PhysRevLett.100.136102} {\bibfield  {journal} {\bibinfo
  {journal} {Physical Review Letters}\ }\textbf {\bibinfo {volume} {100}},\
  \bibinfo {pages} {136102} (\bibinfo {year} {2008})}\BibitemShut {NoStop}%
\bibitem [{\citenamefont {Kolluri}\ and\ \citenamefont
  {Demkowicz}(2010)}]{Kolluri2010}%
  \BibitemOpen
  \bibfield  {author} {\bibinfo {author} {\bibfnamefont {K.}~\bibnamefont
  {Kolluri}}\ and\ \bibinfo {author} {\bibfnamefont {M.~J.}\ \bibnamefont
  {Demkowicz}},\ }\href {\doibase 10.1103/PhysRevB.82.193404} {\bibfield
  {journal} {\bibinfo  {journal} {Physical Review B}\ }\textbf {\bibinfo
  {volume} {82}},\ \bibinfo {pages} {193404} (\bibinfo {year}
  {2010})}\BibitemShut {NoStop}%
\bibitem [{\citenamefont {Kolluri}\ and\ \citenamefont
  {Demkowicz}(2012)}]{Kolluri2012}%
  \BibitemOpen
  \bibfield  {author} {\bibinfo {author} {\bibfnamefont {K.}~\bibnamefont
  {Kolluri}}\ and\ \bibinfo {author} {\bibfnamefont {M.~J.}\ \bibnamefont
  {Demkowicz}},\ }\href {\doibase 10.1103/PhysRevB.85.205416} {\bibfield
  {journal} {\bibinfo  {journal} {Physical Review B}\ }\textbf {\bibinfo
  {volume} {85}},\ \bibinfo {pages} {205416} (\bibinfo {year}
  {2012})}\BibitemShut {NoStop}%
\bibitem [{\citenamefont {Liu}\ \emph {et~al.}(2012)\citenamefont {Liu},
  \citenamefont {Uberuaga}, \citenamefont {Demkowicz}, \citenamefont {Germann},
  \citenamefont {Misra},\ and\ \citenamefont {Nastasi}}]{Liu2012}%
  \BibitemOpen
  \bibfield  {author} {\bibinfo {author} {\bibfnamefont {X.-Y.}\ \bibnamefont
  {Liu}}, \bibinfo {author} {\bibfnamefont {B.~P.}\ \bibnamefont {Uberuaga}},
  \bibinfo {author} {\bibfnamefont {M.~J.}\ \bibnamefont {Demkowicz}}, \bibinfo
  {author} {\bibfnamefont {T.~C.}\ \bibnamefont {Germann}}, \bibinfo {author}
  {\bibfnamefont {A.}~\bibnamefont {Misra}}, \ and\ \bibinfo {author}
  {\bibfnamefont {M.}~\bibnamefont {Nastasi}},\ }\href {\doibase
  10.1103/PhysRevB.85.012103} {\bibfield  {journal} {\bibinfo  {journal}
  {Physical Review B}\ }\textbf {\bibinfo {volume} {85}},\ \bibinfo {pages}
  {012103} (\bibinfo {year} {2012})}\BibitemShut {NoStop}%
\bibitem [{\citenamefont {Daw}\ and\ \citenamefont {Baskes}(1983)}]{Daw1983}%
  \BibitemOpen
  \bibfield  {author} {\bibinfo {author} {\bibfnamefont {M.S.}~\bibnamefont
  {Daw}}\ and\ \bibinfo {author} {\bibfnamefont {M.I.}~\bibnamefont {Baskes}},\
  }\href {\doibase 10.1103/PhysRevLett.50.1285} {\bibfield  {journal} {\bibinfo
   {journal} {Physical Review Letters}\ }\textbf {\bibinfo {volume} {50}},\
  \bibinfo {pages} {1285} (\bibinfo {year} {1983})}\BibitemShut {NoStop}%
\bibitem [{\citenamefont {Daw}\ and\ \citenamefont {Baskes}(1984)}]{Daw1984}%
  \BibitemOpen
  \bibfield  {author} {\bibinfo {author} {\bibfnamefont {M.S.}~\bibnamefont
  {Daw}}\ and\ \bibinfo {author} {\bibfnamefont {M.I.}~\bibnamefont {Baskes}},\
  }\href {\doibase 10.1103/PhysRevB.29.6443} {\bibfield  {journal} {\bibinfo
  {journal} {Physical Review B}\ }\textbf {\bibinfo {volume} {29}},\ \bibinfo
  {pages} {6443} (\bibinfo {year} {1984})}\BibitemShut {NoStop}%
\bibitem [{\citenamefont {Demkowicz}\ and\ \citenamefont
  {Hoagland}(2009)}]{Demkowicz2009}%
  \BibitemOpen
  \bibfield  {author} {\bibinfo {author} {\bibfnamefont {M.~J.}\ \bibnamefont
  {Demkowicz}}\ and\ \bibinfo {author} {\bibfnamefont {R.~G.}\ \bibnamefont
  {Hoagland}},\ }\href {\doibase 10.1142/S1758825109000216} {\bibfield
  {journal} {\bibinfo  {journal} {International Journal of Applied Mechanics}\
  }\textbf {\bibinfo {volume} {1}},\ \bibinfo {pages} {421} (\bibinfo {year}
  {2009})}\BibitemShut {NoStop}%
\bibitem [{\citenamefont {Zhang}\ \emph {et~al.}(2013)\citenamefont {Zhang},
  \citenamefont {Martinez}, \citenamefont {Caro}, \citenamefont {Liu},\ and\
  \citenamefont {Demkowicz}}]{Zhang2013}%
  \BibitemOpen
  \bibfield  {author} {\bibinfo {author} {\bibfnamefont {L.}~\bibnamefont
  {Zhang}}, \bibinfo {author} {\bibfnamefont {E.}~\bibnamefont {Martinez}},
  \bibinfo {author} {\bibfnamefont {A.}~\bibnamefont {Caro}}, \bibinfo {author}
  {\bibfnamefont {X.-Y.}\ \bibnamefont {Liu}}, \ and\ \bibinfo {author}
  {\bibfnamefont {M.~J.}\ \bibnamefont {Demkowicz}},\ }\href {\doibase
  10.1088/0965-0393/21/2/025005} {\bibfield  {journal} {\bibinfo  {journal}
  {Modelling and Simulation in Materials Science and Engineering}\ }\textbf
  {\bibinfo {volume} {21}},\ \bibinfo {pages} {025005} (\bibinfo {year}
  {2013})}\BibitemShut {NoStop}%
\bibitem [{\citenamefont {Kresse}\ and\ \citenamefont
  {Hafner}(1993)}]{Kresse1993}%
  \BibitemOpen
  \bibfield  {author} {\bibinfo {author} {\bibfnamefont {G.}~\bibnamefont
  {Kresse}}\ and\ \bibinfo {author} {\bibfnamefont {J.}~\bibnamefont
  {Hafner}},\ }\href@noop {} {\bibfield  {journal} {\bibinfo  {journal}
  {Physical Review B}\ }\textbf {\bibinfo {volume} {47}},\ \bibinfo {pages} {558} (\bibinfo {year}
  {1993})}\BibitemShut {NoStop}%
\bibitem [{\citenamefont {Kresse}\ and\ \citenamefont
  {Hafner}(1994)}]{Kresse1994a}%
  \BibitemOpen
  \bibfield  {author} {\bibinfo {author} {\bibfnamefont {G.}~\bibnamefont
  {Kresse}}\ and\ \bibinfo {author} {\bibfnamefont {J.}~\bibnamefont
  {Hafner}},\ }\href@noop {} {\bibfield  {journal} {\bibinfo  {journal} {Phys.
  Rev. B}\ }\textbf {\bibinfo {volume} {49}},\ \bibinfo {pages} {14251}
  (\bibinfo {year} {1994})}\BibitemShut {NoStop}%
\bibitem [{\citenamefont {Kresse}\ and\ \citenamefont
  {Furthmuller}(1996)}]{Kresse1996}%
  \BibitemOpen
  \bibfield  {author} {\bibinfo {author} {\bibfnamefont {G.}~\bibnamefont
  {Kresse}}\ and\ \bibinfo {author} {\bibfnamefont {J.}~\bibnamefont
  {Furthmuller}},\ }\href {\doibase 10.1016/0927-0256(96)00008-0} {\bibfield
  {journal} {\bibinfo  {journal} {Comp. Mater. Sci.}\ }\textbf {\bibinfo
  {volume} {6}},\ \bibinfo {pages} {15} (\bibinfo {year} {1996})}\BibitemShut
  {NoStop}%
\bibitem [{\citenamefont {Kresse}\ and\ \citenamefont
  {Furthm\"{u}ller}(1996)}]{Kresse1996a}%
  \BibitemOpen
  \bibfield  {author} {\bibinfo {author} {\bibfnamefont {G.}~\bibnamefont
  {Kresse}}\ and\ \bibinfo {author} {\bibfnamefont {J.}~\bibnamefont
  {Furthm\"{u}ller}},\ }\href@noop {} {\bibfield  {journal} {\bibinfo
  {journal} {Physical review. B, Condensed matter}\ }\textbf {\bibinfo {volume}
  {54}},\ \bibinfo {pages} {11169} (\bibinfo {year} {1996})}\BibitemShut
  {NoStop}%
\bibitem [{\citenamefont {Kresse}\ and\ \citenamefont
  {Joubert}(1999)}]{Kresse1999}%
  \BibitemOpen
  \bibfield  {author} {\bibinfo {author} {\bibfnamefont {G.}~\bibnamefont
  {Kresse}}\ and\ \bibinfo {author} {\bibfnamefont {D.}~\bibnamefont
  {Joubert}},\ }\href {\doibase 10.1103/PhysRevB.59.1758} {\bibfield  {journal}
  {\bibinfo  {journal} {Phys. Rev. B}\ }\textbf {\bibinfo {volume} {59}},\
  \bibinfo {pages} {1758} (\bibinfo {year} {1999})}\BibitemShut {NoStop}%
\bibitem [{\citenamefont {Perdew}\ \emph {et~al.}(1996)\citenamefont {Perdew},
  \citenamefont {Burke},\ and\ \citenamefont {Ernzerhof}}]{Perdew1996}%
  \BibitemOpen
  \bibfield  {author} {\bibinfo {author} {\bibfnamefont {J.~P.~J.}\
  \bibnamefont {Perdew}}, \bibinfo {author} {\bibfnamefont {K.}~\bibnamefont
  {Burke}}, \ and\ \bibinfo {author} {\bibfnamefont {M.}~\bibnamefont
  {Ernzerhof}},\ }\href {\doibase 10.1103/PhysRevLett.77.3865} {\bibfield
  {journal} {\bibinfo  {journal} {Phys. Rev. Lett.}\ }\textbf {\bibinfo
  {volume} {77}},\ \bibinfo {pages} {3865} (\bibinfo {year}
  {1996})}\BibitemShut {NoStop}%
\bibitem [{\citenamefont {Perdew}\ \emph {et~al.}(1997)\citenamefont {Perdew},
  \citenamefont {Burke},\ and\ \citenamefont {Ernzerhof}}]{Perdew1997}%
  \BibitemOpen
  \bibfield  {author} {\bibinfo {author} {\bibfnamefont {J.~P.}\ \bibnamefont
  {Perdew}}, \bibinfo {author} {\bibfnamefont {K.}~\bibnamefont {Burke}}, \
  and\ \bibinfo {author} {\bibfnamefont {M.}~\bibnamefont {Ernzerhof}},\ }\href
  {\doibase 10.1103/PhysRevLett.78.1396} {\bibfield  {journal} {\bibinfo
  {journal} {Phys. Rev. Lett.}\ }\textbf {\bibinfo {volume} {78}},\ \bibinfo
  {pages} {1396} (\bibinfo {year} {1997})}\BibitemShut {NoStop}%
\bibitem [{\citenamefont {Methfessel}\ and\ \citenamefont
  {Paxton}(1989)}]{Methfessel1989}%
  \BibitemOpen
  \bibfield  {author} {\bibinfo {author} {\bibfnamefont {M.}~\bibnamefont
  {Methfessel}}\ and\ \bibinfo {author} {\bibfnamefont {A.~T.}\ \bibnamefont
  {Paxton}},\ }\href@noop {} {\bibfield  {journal} {\bibinfo  {journal}
  {Physical Review B}\ }\textbf {\bibinfo {volume} {40}},\ \bibinfo {pages}
  {3616} (\bibinfo {year} {1989})}\BibitemShut {NoStop}%
\bibitem [{\citenamefont {Plimpton}(1995)}]{Plimpton19951}%
  \BibitemOpen
  \bibfield  {author} {\bibinfo {author} {\bibfnamefont {S.}~\bibnamefont
  {Plimpton}},\ }\href {\doibase http://dx.doi.org/10.1006/jcph.1995.1039}
  {\bibfield  {journal} {\bibinfo  {journal} {Journal of Computational
  Physics}\ }\textbf {\bibinfo {volume} {117}},\ \bibinfo {pages} {1 }
  (\bibinfo {year} {1995})}\BibitemShut {NoStop}%
\bibitem [{\citenamefont {Anderson}\ \emph {et~al.}(2003)\citenamefont
  {Anderson}, \citenamefont {Bingert}, \citenamefont {Misra},\ and\
  \citenamefont {Hirth}}]{Anderson2003}%
  \BibitemOpen
  \bibfield  {author} {\bibinfo {author} {\bibfnamefont {P.~M.}\ \bibnamefont
  {Anderson}}, \bibinfo {author} {\bibfnamefont {J.~F.}\ \bibnamefont
  {Bingert}}, \bibinfo {author} {\bibfnamefont {A.}~\bibnamefont {Misra}}, \
  and\ \bibinfo {author} {\bibfnamefont {J.~P.}\ \bibnamefont {Hirth}},\ }\href
  {\doibase 10.1016/S1359-6454(03)00428-2} {\bibfield  {journal} {\bibinfo
  {journal} {Acta Materialia}\ }\textbf {\bibinfo {volume} {51}},\ \bibinfo
  {pages} {6059} (\bibinfo {year} {2003})}\BibitemShut {NoStop}%
\bibitem [{\citenamefont {Demkowicz}\ and\ \citenamefont
  {Hoagland}(2008)}]{Demkowicz2008}%
  \BibitemOpen
  \bibfield  {author} {\bibinfo {author} {\bibfnamefont {M.~J.}\ \bibnamefont
  {Demkowicz}}\ and\ \bibinfo {author} {\bibfnamefont {R.}~\bibnamefont
  {Hoagland}},\ }\href {\doibase 10.1016/j.jnucmat.2007.02.001} {\bibfield
  {journal} {\bibinfo  {journal} {Journal of Nuclear Materials}\ }\textbf
  {\bibinfo {volume} {372}},\ \bibinfo {pages} {45} (\bibinfo {year}
  {2008})}\BibitemShut {NoStop}%
\bibitem [{\citenamefont {Liu}\ \emph {et~al.}(2010)\citenamefont {Liu},
  \citenamefont {Hoagland}, \citenamefont {Wang}, \citenamefont {Germann},\
  and\ \citenamefont {Misra}}]{Liu2010}%
  \BibitemOpen
  \bibfield  {author} {\bibinfo {author} {\bibfnamefont {X.-Y.}\ \bibnamefont
  {Liu}}, \bibinfo {author} {\bibfnamefont {R.}~\bibnamefont {Hoagland}},
  \bibinfo {author} {\bibfnamefont {J.}~\bibnamefont {Wang}}, \bibinfo {author}
  {\bibfnamefont {T.}~\bibnamefont {Germann}}, \ and\ \bibinfo {author}
  {\bibfnamefont {A.}~\bibnamefont {Misra}},\ }\href {\doibase
  10.1016/j.actamat.2010.05.008} {\bibfield  {journal} {\bibinfo  {journal}
  {Acta Materialia}\ }\textbf {\bibinfo {volume} {58}},\ \bibinfo {pages}
  {4549} (\bibinfo {year} {2010})}\BibitemShut {NoStop}%
\bibitem [{\citenamefont {Kolluri}\ \emph {et~al.}(2013)\citenamefont
  {Kolluri}, \citenamefont {Demkowicz}, \citenamefont {Hoagland},\ and\
  \citenamefont {Liu}}]{Kolluri2013}%
  \BibitemOpen
  \bibfield  {author} {\bibinfo {author} {\bibfnamefont {K.}~\bibnamefont
  {Kolluri}}, \bibinfo {author} {\bibfnamefont {M.~J.}\ \bibnamefont
  {Demkowicz}}, \bibinfo {author} {\bibfnamefont {R.~G.}\ \bibnamefont
  {Hoagland}}, \ and\ \bibinfo {author} {\bibfnamefont {X.-Y.}\ \bibnamefont
  {Liu}},\ }\href {\doibase 10.1007/s11837-012-0540-9} {\bibfield  {journal}
  {\bibinfo  {journal} {JOM}\ }\textbf {\bibinfo {volume} {65}},\ \bibinfo
  {pages} {374} (\bibinfo {year} {2013})}\BibitemShut {NoStop}%
\bibitem [{\citenamefont {Ercolessi}\ and\ \citenamefont
  {Adams}(1994)}]{Ercolessi1994}%
  \BibitemOpen
  \bibfield  {author} {\bibinfo {author} {\bibfnamefont {F.}~\bibnamefont
  {Ercolessi}}\ and\ \bibinfo {author} {\bibfnamefont {J.~B.}\ \bibnamefont
  {Adams}},\ }\href {\doibase 10.1209/0295-5075/26/8/005} {\bibfield  {journal}
  {\bibinfo  {journal} {Europhysics Letters (EPL)}\ }\textbf {\bibinfo {volume}
  {26}},\ \bibinfo {pages} {583} (\bibinfo {year} {1994})}\BibitemShut
  {NoStop}%
\bibitem [{\citenamefont {Fellinger}\ \emph {et~al.}(2010)\citenamefont
  {Fellinger}, \citenamefont {Park},\ and\ \citenamefont
  {Wilkins}}]{Fellinger2010}%
  \BibitemOpen
  \bibfield  {author} {\bibinfo {author} {\bibfnamefont {M.~R.}\ \bibnamefont
  {Fellinger}}, \bibinfo {author} {\bibfnamefont {H.}~\bibnamefont {Park}}, \
  and\ \bibinfo {author} {\bibfnamefont {J.~W.}\ \bibnamefont {Wilkins}},\
  }\href {\doibase 10.1103/PhysRevB.81.144119} {\bibfield  {journal} {\bibinfo
  {journal} {Physical Review B}\ }\textbf {\bibinfo {volume} {81}},\ \bibinfo
  {pages} {144119} (\bibinfo {year} {2010})}\BibitemShut {NoStop}%
\end{thebibliography}%

\end{document}